\documentclass[
reprint,
 amsmath,amssymb,
 aps,
prb,
longbibliography
]{revtex4-1}

\usepackage{graphicx}
\usepackage{dcolumn}
\usepackage{bm}
\usepackage{color}
\usepackage{braket}
\usepackage{multirow}
\usepackage[colorlinks,linkcolor=blue, urlcolor=blue,citecolor=blue,bookmarks=fales]{hyperref}

\begin{document}
\title{Orbital-dependent quasiparticle scattering interference in 3R-NbS$_2$}

\author{T. Machida}
\email{tadashi.machida@riken.jp}
\affiliation{RIKEN Center for Emergent Matter Science, Wako, Saitama 351-0198, Japan}

\author{Y. Kohsaka}
\affiliation{RIKEN Center for Emergent Matter Science, Wako, Saitama 351-0198, Japan}

\author{R. Suzuki}
\affiliation{Department of Applied Physics and Quantum Phase Electronics Center (QPEC), The University of Tokyo, Hongo, Bunkyo-ku, Tokyo 113-8656, Japan}

\author{K. Iwaya}
\affiliation{RIKEN Center for Emergent Matter Science, Wako, Saitama 351-0198, Japan}

\author{M. Ochi}
\affiliation{Department of Physics, Osaka University, Toyonaka, Osaka 560-0043, Japan}

\author{R. Arita}
\affiliation{RIKEN Center for Emergent Matter Science, Wako, Saitama 351-0198, Japan}

\author{T. Hanaguri}
\affiliation{RIKEN Center for Emergent Matter Science, Wako, Saitama 351-0198, Japan}

\author{Y. Iwasa}
\affiliation{RIKEN Center for Emergent Matter Science, Wako, Saitama 351-0198, Japan}
\affiliation{Department of Applied Physics and Quantum Phase Electronics Center (QPEC), The University of Tokyo, Hongo, Bunkyo-ku, Tokyo 113-8656, Japan}

\begin{abstract}
A valley degree of freedom (DOF) in transition metal dichalcogenides with broken inversion symmetry can be controlled through spin and orbital DOFs owing to their valley-contrasting characters.
Another important aspect of the spin and orbital DOFs is that they affect quasiparticle scattering processes that govern the valley lifetime.
Here we combine quasiparticle-interference (QPI) imaging experiments and theoretical simulations to study the roles of the spin and orbital DOFs in 3R-NbS$_2$.
We find that the QPI signal arising from an inter-valley scattering is noticeably weaker than that caused by an intra-valley scattering.
We show that this behavior is predominantly associated with the orbital DOF, signifying the different spin and orbital structures of spin-split bands at each valley.
These findings provide important insights into understanding the valley-related transport properties.
\end{abstract}

\maketitle

\section{Introduction}
Degrees of freedom (DOFs) of electrons in solids are the basis of information processing, such as charge and spin DOFs in electronics and spintronics, respectively.
Valleys in momentum space, namely multiple local extrema in an electronic band structure, have been proposed as an emergent DOF, being expected to open novel applications (valleytronics)~\cite{Gunawan2006PRL,Rycerz2007NP, Xiao2007PRL}.
Systems with a honeycomb-like lattice structure, such as graphene and transition metal dichalcogenides  (TMDCs), can possess two energetically degenerated valleys at the corners of the Brillouin zone ($\pm$K points) and offer a facilitating stage for valleytronics~\cite{Rycerz2007NP,Neto2009RMP,Xiao2007PRL,Yao2008PRB,Zhang2011PRL,Xiao2010RMP,Xiao2012PRL,Yu2015PRL,Xu2014NP,Ando2005JPSJ}.

Manipulation of the valley DOF is a prerequisite for valleytronics and can be achieved through valley-contrasting attributes.
A noticeable example is the valley Hall effect caused by the valley-contrasting Berry curvature~\cite{Xiao2007PRL,Yao2008PRB,Zhang2011PRL,Xiao2010RMP, Xiao2012PRL,Yu2015PRL,Xu2014NP,Mak2014Science,Shimazaki2015NP,JLee2016NNano}.
Monolayer TMDCs with broken inversion symmetry offer alternative ways to manipulate the valley DOF through the spin and orbital DOFs.
The broken inversion symmetry and strong spin-orbit coupling create not only valley-spin coupling characterized by opposite spin polarizations at $\pm$ K valleys but also valley-orbit coupling signified by valley-contrasting orbital characters $d_{x^2-y^2} \pm id_{xy}$ at $\pm$ K valleys~\cite{Xiao2012PRL,Yu2015PRL,Xu2014NP}.
These couplings allow us to control the valley polarization by an external magnetic field ~\cite{Srivastava2015NP,Aivazian2015NP,Li2014PRL,MacNeill2015PRL} and by circularly polarized light~\cite{Xiao2012PRL,Yu2015PRL,Xu2014NP, Cao2012NC,Zeng2012NNano,Mak2012NNano}, for example.

The spin and orbital DOFs also affect quasiparticle scattering processes that govern transport properties including a valley lifetime.
Therefore it is crucial to investigate the roles of the spin and orbital DOFs experimentally.
Spin- and angle-resolved photoemission spectroscopy is a powerful tool that can directly access to the spin structures in momentum space.
A clear signature of the valley-spin coupling has been first demonstrated in 3R-type bulk MoS$_2$~\cite{Suzuki2014NNano}.
Quasiparticle interference (QPI) patterns observed by spectroscopic-imaging scanning tunneling microscopy (SI-STM) are also sensitive to the spin and orbital DOFs and have provided information on both spin and orbital structures in momentum space~\cite{Roushan2009Nature,Beidenkopf2011NP,Okada2011PRL,Brihuega2008PRL,Mallet2012PRB,Huang2014PRB,Singh2015SA,Zeljkovic2014NP}.
QPI patterns are generated by an interference between two quantum states on the same constant-energy contours in momentum space.
If these states are orthogonal, they can not interfere with each other and no QPI signal is generated at the scattering vector $\bm{q}$ connecting these two states.
Since the orthogonality is related to both spin and orbital DOFs, one can, in principle, infer the roles played by the spin and orbital DOFs through the selection rule of QPI.
In monolayer WSe$_2$, a QPI signal corresponding to the inter-valley scattering is suppressed or absent~\cite{HLiu2015NC,Yankowitz2015PRL}.
This has been attributed to the opposite spin orientations at $\pm$~K valleys.
However, since the orbital DOF is also active in the TMDCs, distinguishing the role played by each DOF requires careful considerations.

In this article, we investigate QPI patterns of 3R-NbS$_2$ and analyzed the data using the theory based on the $T$-matrix formalism where both spin and orbital DOFs are treated on an equal footing.
Bulk 3R-NbS$_2$ was chosen for the present purpose, because of its broken inversion symmetry with conducting states. 
We find that QPI signals of the inter-valley scatterings are much weaker than those caused by the intra-valley scatterings.
We have succeeded in reproducing this behavior by theoretical simulations and reveal that the selection rule associated with the orbital DOF governs the QPI patterns.
We also point out that different spin and orbital structures between the spin-split bands at each valley are important for the inter-valley scattering, providing a clue for understanding valley-related transport properties.

\section{Methods}
Single crystals of 3R-NbS$_2$ were grown by chemical vapor transport technique. A mixture of Nb, S and I$_2$ was sealed in a quartz tube and placed in a two-zone horizontal temperature gradient furnace: the higher-temperature side was kept at 1100 $^{\circ}$C and lower side at 900 $^{\circ}$C for 12 days.
We confirmed that the polytype of the grown crystals is 3R type by powder X-ray measurements.
The STM measurements were done at a temperature of 4.6~K using a commercial low-temperature ultra-high-vacuum STM (Unisoku USM-1300) modified by ourselves~\cite{Hanaguri2006JPhys}.
The samples were cleaved under ultra-high-vacuum ($< 10^{-9}$ Torr) at 77~K and immediately transferred to the STM head that was kept below $\sim$10~K.
All of the spectroscopic measurements were done by the standard lock-in technique.

We also performed first-principles band structure calculations using the Perdew-Burke-Ernzerhof parameterization of the generalized gradient approximation \cite{Perdew1996} and the full-potential (linearized) augmented plane-wave method with an inclusion of the spin-orbit coupling as implemented in the \textsc{wien2k} code \cite{Blaha2001}. We used the experimental crystal structure shown in Ref. \onlinecite{expt_strct}. We set the muffin-tin radii of Nb and S atoms to 2.48 and 2.14 Bohr, respectively, and adopted $RK_{\mathrm{max}}=7.00$.

\section{Results and Discussion}
First we introduce basic electronic structures of 3R-NbS$_2$ obtained by first-principles calculations.
The crystal structure of 3R-NbS$_2$ globally breaks inversion symmetry as shown in Fig.~1(a).
As a result, the band structure is similar to that of monolayer TMDCs, which is characterized by the spin-split bands at $\pm$K points that constitute valleys [Fig.~1(b)].
Valley-contrasting character manifests itself in the spin-split bands at $\pm$K points; at +K point, the spin-up band is higher than the spin-down band whereas the situation is opposite at -K point [Fig.~1(d)].
The energy splitting at $\pm$K points is calculated to be as large as 120~meV.
The Fermi level of 3R-NbS$_2$ lies deep inside the spin-split bands, which brings about metallic conduction making SI-STM experiments possible.
As schematically shown in Fig.~1(d), $\Gamma_1$ and $\Gamma_2$ bands at the center of the Brillouin zone are three dimensional in nature and thus the QPI signals related to these bands may be broadened.
The valley bands at $\pm$K are almost two dimensional, being expected to govern the QPI signals.

Figure~2(a) depicts a typical topographic image that shows a triangular lattice with a lattice constant of 3.4~\AA, which is consistent with the bulk $a$-axis constant.
From the observed lattice, we determine the crystallographic orientation, as indicated by the black arrows in Fig.~2(a).
Although both of S and Nb sublattices have the same lattice constant, we infer that the former is imaged because the cleaving occurs between the neighboring S atoms bonded by the weak van der Waals interaction.
There are many defects imaged as protrusions ($\sim$ 8 \% of Nb atoms) and depressions ($\sim$ 4\% of Nb atoms).
Both of them occupy one of the hollow sites as marked by the red and yellow crosses in the inset of Fig.~2(a).
As shown in Fig.~1(a), there are two different hollow sites that correspond to the Nb site of the topmost NbS$_2$ layer and the S site of the second topmost NbS$_2$ layer.
Given that few defects are observed in the imaged topmost S sublattice, we postulate that the defects are not located at the S site but at the Nb site.
Such defects should modify low-energy states consisting of Nb $d$-orbitals and may work as quasiparticle scatters that generate QPI patterns.

Figure~2(b) shows the measured tunneling spectrum $dI/dV$ reflecting the density of states (DOS).
Here, $I$ and $V$ denote the tunneling current and the sample bias voltage, respectively.
The calculated DOS below $E_{\rm{F}}$ possesses several peaks or humps that are not clearly observed in the tunneling spectrum as shown in Fig.~2(c). However, it qualitatively capture the tunneling spectrum  above $E_{\rm{F}}$: the gradual reduction of the DOS above $\sim$ +300 meV and the sudden growth of the DOS above $\sim$ +2.1 eV that corresponds to the bottom of the band lying above +2.1 eV [Fig.~1(b)].
The valley bands lie in an energy range roughly from $E_{\rm{F}}$ to +1~eV as marked by the black box in Fig.~1(b).
Therefore, we performed SI-STM experiments in this energy range to image QPI patterns.

Figures~3(a) to~3(e) show series of differential conductance maps $g(\bm{r}, E) \equiv dI(\bm{r})/dV|_{V=E/e}$ taken in the same field-of-view of Fig.~2(a).
Here, $\bm{r}$ denotes position at the surface, $E$ is the energy and $e$ is the elementary charge.
We perform Fourier transformation of $g(\bm{r}, E)$ to obtain the $\bm{q}$-resolved conductance map $g_q(\bm{q}, E)$ in which QPI signals show up as local maxima.
To emphasize the possibly small QPI signals, we take a second derivative of $g_q(\bm{q}, E)$ with respect to $E$ as shown in Figs.~3(k)-3(o), where the white colored intensities indicate the center of the QPI intensities.
Features appearing near the origin of $g_q(\bm{q}, E)$ correspond to the small $\bm{q}$ scatterings including the intra-valley scattering.
There are three ring-like features that spread out with decreasing energy.
[Halves of them are marked in green, cyan, and orange arcs in Fig.~3(k)-3(o).]
We examine the energy-dependent line profile along the red line shown in Fig.~3(o) to identify the origins of these QPI signals [Fig.~4(c)].
Three hole-like branches guided by green, cyan, and orange lines correspond to the three rings in Figs.~3(k)-3(o).
There are two local maxima in intensity at $\bm{q} = 0$ around +750~meV and +600~meV as pointed by the two arrows in Fig.~4(c).
Because the energy difference between them $\sim$150 meV is close to the calculated energy splitting of spin-split bands at $\pm$K points $\sim$120 meV, we assign that they are the band edges of upper and lower spin-split valley bands.
(The energy difference between the observed and the calculated spin split bands is about 65 meV.)
This means that the middle branch (cyan) in Fig.~4(c) stems from the intra-valley scattering $\bm{q}_{\mathrm{intra}}$ schematically shown in Fig.~1(d).
The uppermost (green) and lowermost (orange) branches may originate from the intra-band scatterings of $\Gamma_1$ and $\Gamma_2$-band, respectively.
Note that the intensity of the uppermost branch is weak and the lowermost branch is rather broad.
These observations may be related to the three dimensionality of the $\Gamma_1$ and $\Gamma_2$ band~\cite{Suzuki2014NNano}.

Besides the intra-valley QPI signal, an inter-valley QPI [denoted as $\bm{q}_{\mathrm{inter}}$ in Fig.~1(d)] signal is expected in the vicinity of the $\bm{q}$ point marked by crosses in Fig.~3(f)-3(o).
At high energies, the inter-valley QPI signal is hardly detected but below about +400~meV, a faint feature emerges as indicated by the red dashed circle in Fig.~3(m).
This feature splits and merges with neighboring signals with decreasing energy.
The characteristics of this inter-valley QPI signal is highlighted in Fig.~4(b) and ~4(d) that show the energy-dependent line profiles of $g_q(\bm{q}, E)$ and $d^2g_q/dE^2(\bm{q}, E)$ along the white line in Fig.~3(o), respectively.
The most noticeable observation is that the inter-valley QPI signal is much weaker than the intra-valley counterpart especially near the band edge energies [Fig.~4(a) and ~4(b)].
Only one branch appears below +400 meV as shown in Fig.~4(d).
In principle, another QPI signal due to the scattering from +K valley to the separate +K valley may show up around the Bragg spot (G) but it was not detected.
The absence of this QPI signal may be due to the finite resolution of the STM tip and/or the finite width of the impurity potential, both of which suppress large-$\bm{q}$ QPI signals.
Although this effect might suppress the inter-valley QPI signals, it seems to be not enough to reproduce the large intensity difference between the intra- and inter-valley QPI signals.
The clear intra-valley QPI signals survive up to $|\bm{q}| \sim$ 1.0 \AA$^{-1}$ (at $E$ = 0 meV) as shown in Fig. 4(a). Even though the $|\bm{q}|$ of the inter-valley QPI ($\sim$ 1.26 \AA$^{-1}$) is only 30\% longer than the value of $|\bm{q}| \sim $ 1.0 \AA$^{-1}$, the actual inter-valley QPI intensity is noticeably weaker or absence compared with the intra-valley QPI signals.
This observation implies that another mechanism is indispensable to explain the suppression of the inter-valley QPI signals.

The observed contrast between intra- and inter-valley scatterings should carry information of the spin and orbital DOFs at the $\pm$K valleys.
To unveil the roles of these DOFs, we performed theoretical QPI simulations in which the spin and orbital DOFs are hypothetically suppressed one by one.
(See Appendix A for details of the QPI simulations.)
Consequently, we are able to compare our experimental data with four different simulated results, each of which is characterized by its own selection rule(s) associated with the active DOFs [Fig.~4(e)-4(l)].
As shown in Figs.~4(e) and 4(f), the result of the full simulation including both spin and orbital selection rules reproduces very well the observed QPI signals of the intra- and inter-valley scatterings in terms of the energy dispersions, the energy-dependent intensities and the contrast between the intra- and inter-valley scatterings.
On the other hand, the simulation that ignores both selection rules [Figs.~4(k) and 4(l)] exhibits that the QPI intensity is almost independent of the energy and the wave vector $\bm{q}$. 
This $E$- and $\bm{q}$-independent behavior of the QPI intensity apparently contradicts with the experimental result.
Namely, the spin and/or orbital DOFs are relevant to the real QPI signals.

We can separate the effects of the spin and orbital DOFs by virtually suppressing one of these DOFs.
As shown in Figs.~4(g) and 4(h), it is clear that the result of the simulation including only the orbital DOF is very similar to the result of the full simulation, as well as to the experimental observation.
By contrast, the simulation including only the spin DOF [Figs.~4(i) and 4(j)] does not account for the $E$- and $\bm{q}$-dependence of the QPI intensities even though it suppress some of the branches appearing in the simulation without both selection rules [Figs.~4(k) and 4(l)].
Therefore, we can safely conclude that the orbital DOF plays a pivotal role in the QPI in 3R-NbS$_2$.

Such contrasting consequences brought by the spin and orbital DOFs can be understood in terms of the different spin and orbital characters of valley bands.
Figure~4(m) depict the calculated spin- and orbital-resolved spectral function.
First we discuss the effect of orbital DOF.
It is clear that the $d_{x^2-y^2} + id_{xy}$ ($d_{x^2-y^2} - id_{xy}$) orbital component is almost completely localized at the +K (-K) valley, representing the valley-contrasting orbital angular momentum.
In such a situation, the intra-valley scattering is possible but the inter-valley scattering is forbidden by the orbital-selection rule; the initial and final states of the scattering are orthogonal with each other in orbital space.
A gradual increase of the inter-valley scattering intensity at low energies is due to the mixed $d_{z^2}$ component that exists equally at both valleys and its spectral weight increases with decreasing energy.

The effect of the spin DOF is different from that of the orbital DOF.
This is because spin orientations of the valley bands are reversed not only between the $\pm$K valleys but also between the spin-split bands at each valley.
(Note that the orbital character is maintained between the upper and lower spin-split bands.)
Near the top of the upper spin-split band, the spin-selection rule prevents the inter-valley scattering, as in the case of the orbital-selection rule, since spin orientations are opposite between the $\pm$K valleys.
However, below the top of the lower spin-split bands, there appears an allowed inter-valley scattering channel with the same spin orientation between the initial and final states, unless other selection rules apply [Fig.~4(n)].
This is the reason why an intense inter-valley branch remains in the simulation including only the spin DOF [Fig.~4(j)].

We schematically summarize our interpretation in Fig.~4(n).
In the intra-valley scattering, there are three possible scattering channels.
One of these three is prevented by the spin-selection rule but other two are allowed.
In the inter-valley scattering, there are four possible scattering channels.
Even though the spin-selection rule allows two of them, the orbital-selection rule prevent all of the channels.
As a results, there appears a large intensity difference between the QPI signals caused by the intra- and inter-valley scatterings.

\section{Conclusion}

We study the QPI patterns in the valleytronics-candidate TMDC compound 3R-NbS$_2$ by using Fourier transformed SI-STM to clarify the roles played by the spin and orbital DOFs.
The observed inter-valley QPI signal is noticeably weaker than the intra-valley QPI signal.
A comparison with the theoretical simulations highlights the pivotal role of the orbital DOF in the observed QPI patterns.
We show that this is related not only to the valley-contrasting characters of the spin and orbital DOFs but also to the difference in the spin and orbital structures at each valley; the orbital characters are maintained between the spin-split valley bands whereas the spin orientations are reversed between them.
Our results clearly indicate that not only the spin DOF but also the orbital DOF should be explicitly taken into account to analyze the quasiparticle scattering processes in TMDCs.
This is important information for many valley-related phenomena, such as a valley lifetime that governs the amplitude of the valley Hall effect.

\section*{Acknowledgments}
This work was supported by JSPS (KAKENHI Grants No. 16K17730 and 2500003).
R. S. was supported by JSPS through a research fellowship for young scientists and by the Leading Graduate Program of Materials Education for future leaders in Research, Industry and Technology (MERIT).

\begin{figure*}[t]
\begin{center}
\includegraphics[width=8cm]{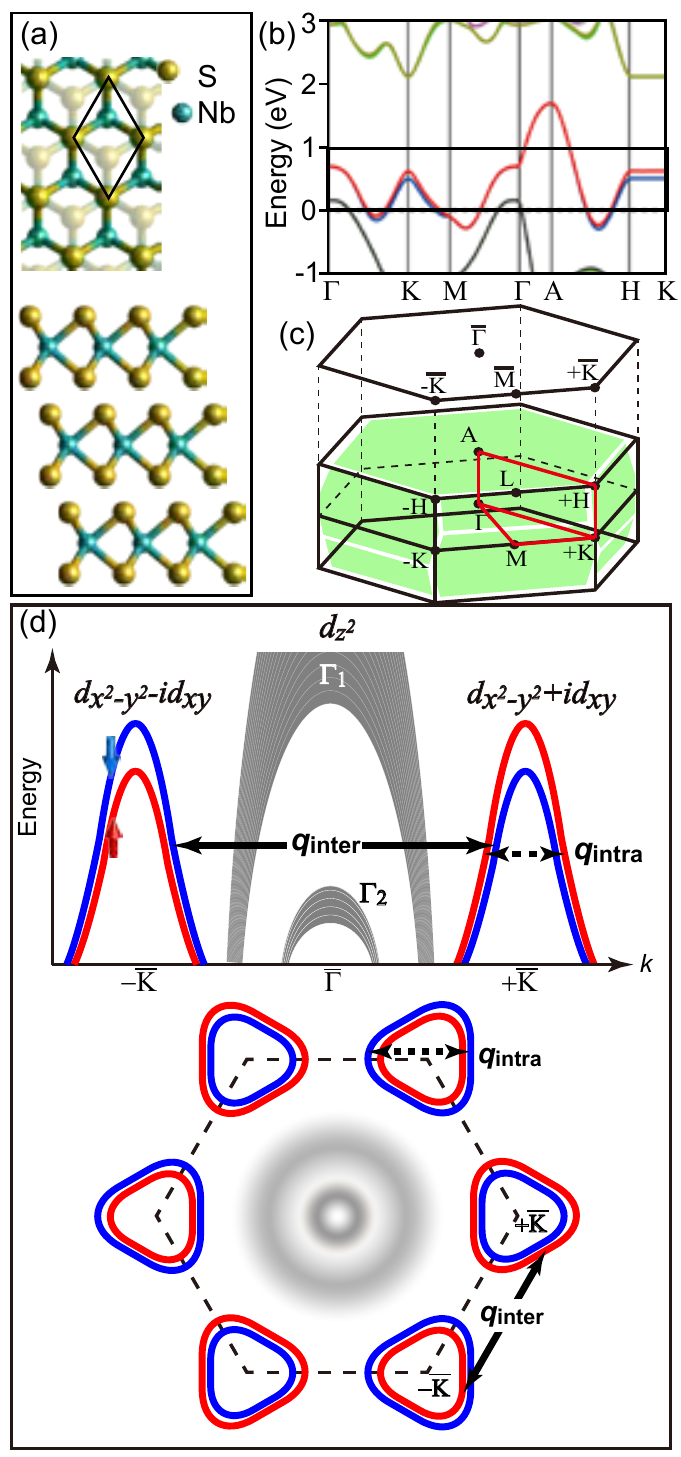}
\end{center}
\caption{
(a) Crystal structure of 3R-NbS$_2$. Yellow and cyan spheres indicate S and Nb ions, respectively.
(b) Band structure of 3R-NbS$_2$ by first-principles calculations (See methods section for detailed procedure) along a pass through the high symmetry points in the first Brillouin zone indicated by the red lines in (c).
A black box denotes an energy range on which we focus in this study.
(c) First BZ of 3R-NbS$_2$. Green shaded areas denote the original BZ of 3R-NbS$_2$. Solid lines represent the conventional BZ for hexagonal unit cell.
The two-dimensional projected BZ is represented by the topmost hexagonal plane containing $\overline{\Gamma}$, $\overline{M}$, and $\pm\overline{K}$.
(d) Schematic illustrations of band structure within the energy range indicated by the black box in (b). 
Bottom panel represents constant energy contours at $\sim$ +100 meV. 
Bands in red and blue represent quasi-2D spin-up and -down valley-bands around $\mathrm{\pm}$K points, respectively. Gray shaded bands denote spin-degenerated three dimensional bands around $\Gamma$, named as $\Gamma_1$- and $\Gamma_2$-band.
Dashed and solid arrows indicate the intra- $\bm{q}_{\rm{intra}}$ and inter-valley $\bm{q}_{\rm{inter}}$ scattering channels, respectively.
}
\end{figure*}

\begin{figure*}[t]
\begin{center}
\includegraphics[width=8cm]{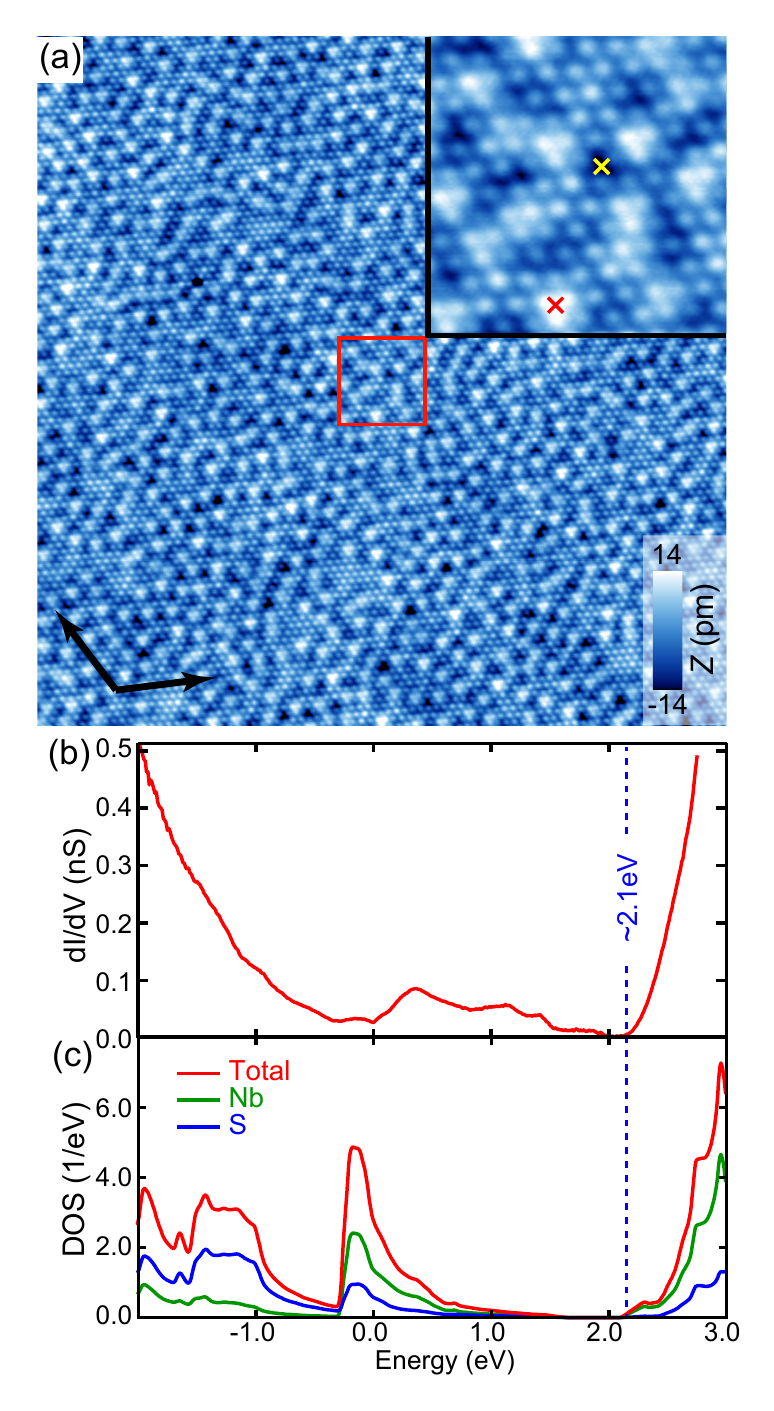}
\end{center}
\caption{
(a) A topographic image taken at $V = $ +1.0 V and $I = 200$ pA on a 300 $\times$ 300 \AA$^2$ field of view.
Inset represents a magnified image of an area marked by a red box in the main image.
Red and yellow crosses indicate examples of triangular protrusions and depressions, respectively.
(b) A typical tunneling spectrum taken by the standard lock-in technique with a modulation amplitude of 20~mV$_{\rm rms}$ at a set point of $V =$ -0.5 V and $I$ = 100 pA.
(c) Density-of-states by first-principles calculations. Red, blue, and green lines denote total DOS, partial DOS of Nb, and of S ions, respectively.
A dashed blue line indicates the energy at which both the tunneling spectrum and the calculated DOS exhibit the rapid enhancement.
}
\end{figure*}

\begin{figure*}[t]
\begin{center}
\includegraphics[width=17cm]{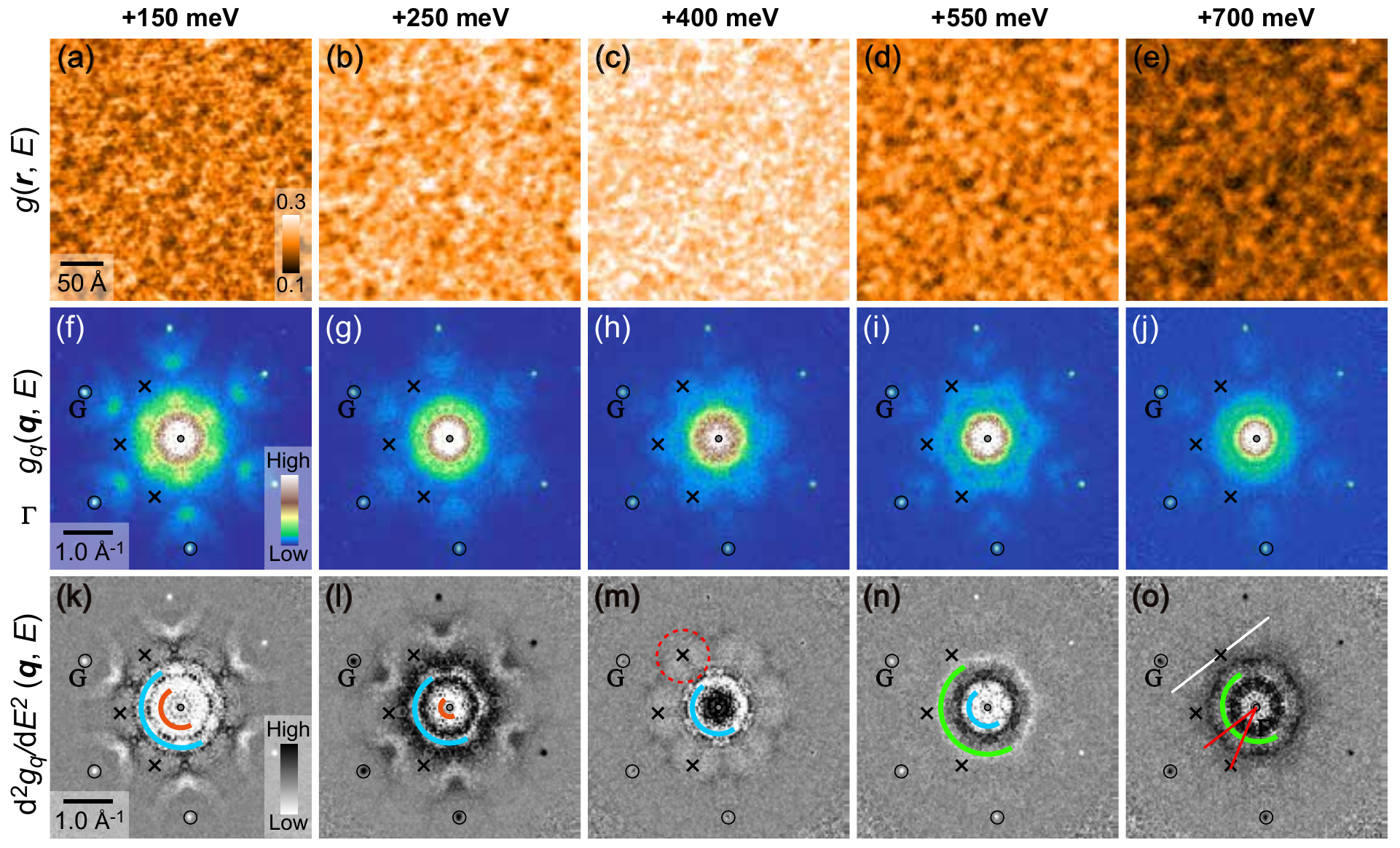}
\end{center}
\caption{
(a) to (e) Differential conductance maps; $g(\bm{r}, E)$ at $E = $ +150, +250, +400, +550, and +700 meV, respectively. These maps are taken by  the standard lock-in technique with a modulation amplitude of 20~mV$_{\rm rms}$ at a set point of $V$ = 1 V and $I$ = 200 pA. 
(f) to (j) Fourier transformed conductance maps;  $g_q(\bm{q}, E)$ at $E = $ +150, +250, +400, +550, and +700 meV, respectively.
(k) to (o) Second derivative of  $g_q(\bm{q}, E)$ at $E = $ +150, +250, +400, +550, and +700 meV, respectively. White colored intensities represent a negative value of the second derivative of $g_q(\bm{q}, E)$ and the center of the QPI intensities.
The green, cyan, and orange arcs represent three intra-band or -valley QPI signals around $\Gamma$.
Red dashed circle denotes inter-valley QPI signal around $\bm{q}$ points marked by black crosses.
}
\end{figure*}

\begin{figure*}[t]
\begin{center}
\includegraphics[width=18cm]{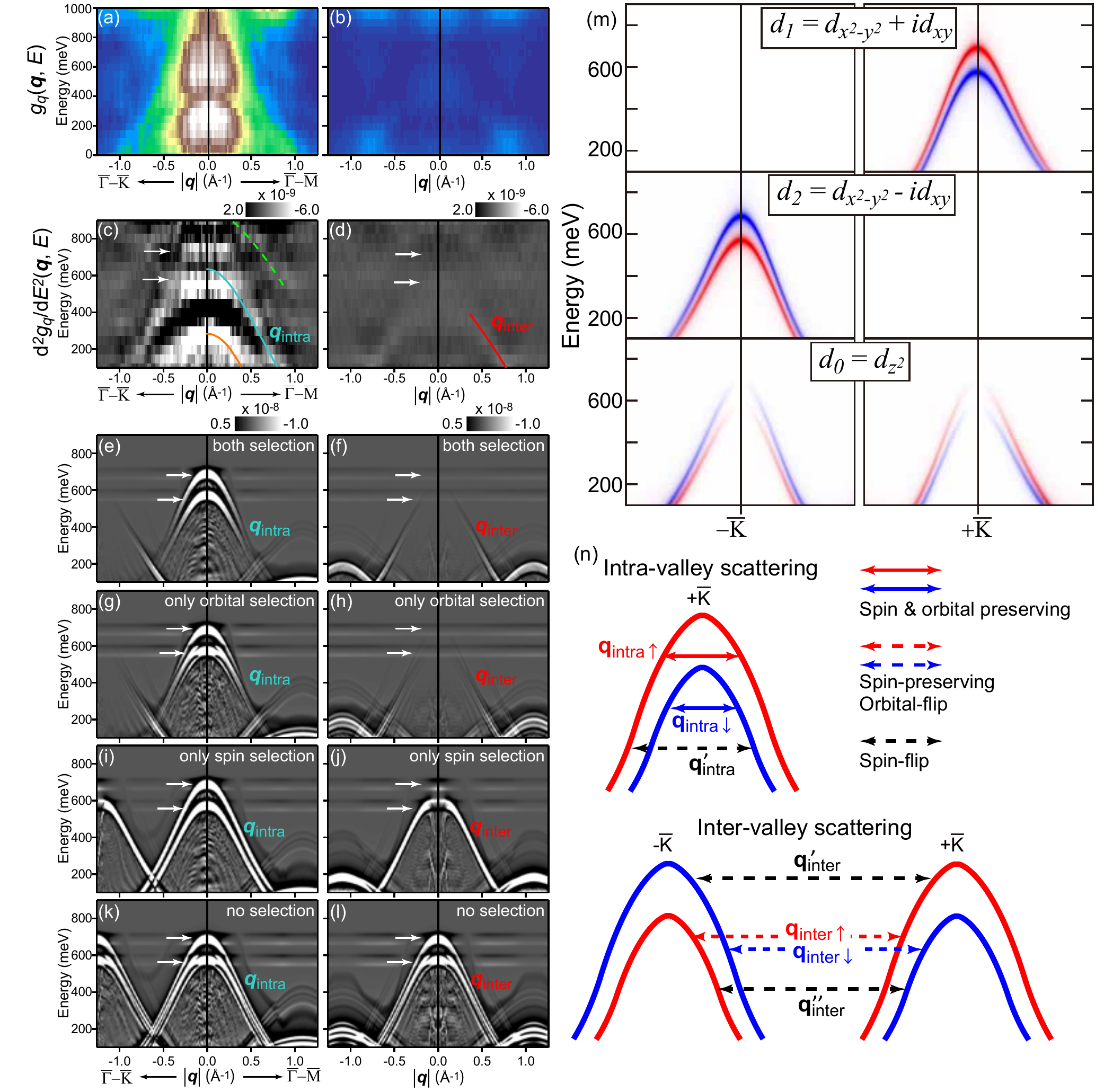}
\end{center}
\caption{
(a) and (b) Energy dependent line profiles of  $g_q(\bm{q}, E)$ from 0 to +1000 meV along the red and white lines in Fig.~3(o), respectively.
(c) and (d) Line profiles of $d^2g_q/dE^2(\bm{q}, E)$ from +100 to +900 meV along the same directions as (a) and (b), respectively. Note that we exclude the data below +100 meV and above +900 meV which contain the end effect of the energy second derivation.
(e) and (f) Simulated line profiles of $d^2g_q/dE^2(\bm{q}, E)$ along the same directions as (a) and (b), respectively.
Both the spin and orbital DOFs are included for the simulation. 
(g) and (h) Same as (e) and (f) but only the orbital DOF is included.
(i) and (j) Same as (e) and (f), but  only the spin DOF is included.
(k) and (l) Same as (e) and (f), but both DOFs are ignored.
White colored intensities in (c) to (l) and track the original QPI dispersions.
Two white arrows in (c)-(l) denote the $E_{\rm{USB}}$ and $E_{\rm{LSB}}$.
(m) Calculated spin- and orbital-resolved spectral weights of the valley bands. 
Top, middle and bottom panels denote $d_1$, $d_2$, and $d_0$ components, respectively.
The spin-up and -down components are shown by red and blue, respectively.
(n) Schematic illustrations of the intra- (upper) and inter-valley (lower) scattering channels. 
Solid arrows in red and blue denote the scatterings conserving both spin and orbital characters between the initial and final states.
Dashed black arrows indicate the scatterings flipping both spin and orbital characters.
Dashed red and blue arrows represent the scatterings conserving the spin but flipping the orbital character.
}
\end{figure*}

\appendix
\section{Tight-binding Hamiltonian and QPI simulations}

For QPI simulations, we build three-orbital third-nearest-neighbor tight-binding Hamiltonian with the same basis set of Ref.~\onlinecite{GBLiu2013PRB},
\begin{equation}
	\label{eq:OriginalBasis}
	\psi^{\dag} = (d^{\dag}_{z^2\uparrow},\ d^{\dag}_{z^2\downarrow},\ d^{\dag}_{x^2-y^2\uparrow},\ d^{\dag}_{x^2-y^2\downarrow},\ d^{\dag}_{xy\uparrow},\ d^{\dag}_{xy\downarrow})^{T}.
\end{equation}
The tight-binding parameters are determined by fitting the first-principles band shown in Fig.~5.
The obtained parameters are summarized in Table I.
\begin{table}[htb]
\caption{Tight binding parameters (in units of eV) determined by fitting the first-principles band structure. $\epsilon_1$ and $\epsilon_2$ denote the on-site energy in $d_{z^{2}}$ and $d_{x^2-y^2}$ or $d_{xy}$ orbitals, respectively. $t$, $r$, and $u$ are nearest neighbor (NN), next nearest neighbor (NNN), third nearest neighbor (TNN) hopping parameters, respectively. $\lambda$ represents the spin-orbit coupling. In this study, we fix the lattice constant $a_0 = 3.33$ \AA\ and added $E_{\mathrm{offset}}$ = 0.065 to the diagonal components of the Hamiltonian, so that the calculated spin-split band matches the experimental one.
}
{\tabcolsep = 1.5mm
  \begin{tabular}{c|ccc}
  \hline
  \hline
    & Orbital & Notation & Energy (eV) \\ \hline
    \multirow{2}{*}{on-site} & $d_{z^2}$ & $\epsilon_1$ & 0.9775 \\
    & $d_{xy}$, $d_{x^2-y^2}$  & $\epsilon_2$ & 2.5438 \\ \hline
    \multirow{6}{*}{NN} & $d_{z^2} \rightarrow d_{z^2}$ & $t_0$ & -0.0928 \\
                                      & $d_{z^2} \rightarrow d_{xy}$ & $t_1$ & 0.3629 \\
                                      & $d_{z^2} \rightarrow d_{x^2-y^2}$ & $t_2$ & 0.3339 \\
                                      & $d_{xy} \rightarrow d_{xy}$ & $t_{11}$ & 0.1057 \\
                                      & $d_{xy} \rightarrow d_{x^2-y^2}$ & $t_{12}$ & 0.2454 \\
                                      & $d_{x^2-y^2} \rightarrow d_{x^2-y^2}$ & $t_{22}$ & 0.0050 \\
  \hline
    \multirow{5}{*}{NNN} & $d_{z^2} \rightarrow d_{z^2}$ & $r_0$ & 0.1124 \\
                                      & $d_{z^2} \rightarrow d_{xy}$ & $r_1$ & -0.1066 \\
                                      & $d_{z^2} \rightarrow d_{x^2-y^2}$ & $r_2$ & 0.1297 \\
                                      & $d_{xy} \rightarrow d_{xy}$ & $r_{11}$ & 0.0191 \\
                                      & $d_{xy} \rightarrow d_{x^2-y^2}$ & $r_{12}$ & -0.0069 \\
  \hline
    \multirow{6}{*}{TNN} & $d_{z^2} \rightarrow d_{z^2}$ & $u_0$ & -0.0649 \\
                                      & $d_{z^2} \rightarrow d_{xy}$ & $u_1$ & 0.0276 \\
                                      & $d_{z^2} \rightarrow d_{x^2-y^2}$ & $u_2$ & -0.0301 \\
                                      & $d_{xy} \rightarrow d_{xy}$ & $u_{11}$ & 0.1572 \\
                                      & $d_{xy} \rightarrow d_{x^2-y^2}$ & $u_{12}$ & -0.0780 \\
                                      & $d_{x^2-y^2} \rightarrow d_{x^2-y^2}$ & $u_{22}$ & -0.0082 \\
  \hline
  SO coupling & - & $\lambda$ & 0.0575 \\
  \hline
  \hline
  \end{tabular}
}
\end{table}

\begin{figure*}[t]
\begin{center}
\includegraphics[width=10cm]{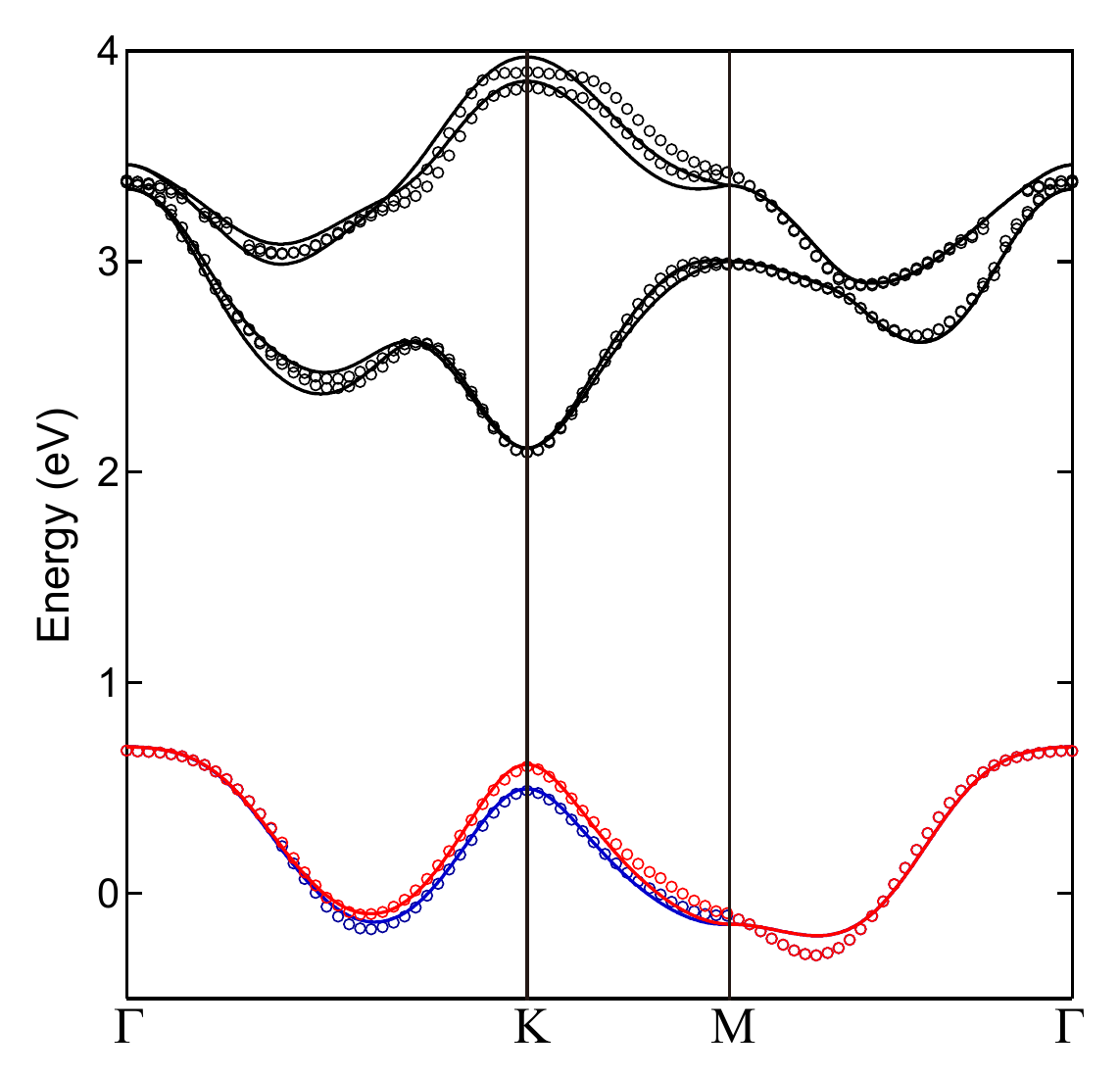}
\end{center}
\caption{
Energy bands of 3R-NbS$_2$. Circles and lines denote the first-principles and the tight binding results, respectively.
The spin-up and -down bands on which we focus in this study are represented by red and blue, respectively.
}
\end{figure*}

To describe more explicitly the valley-orbit coupling, we change the basis set by introducing the unitary transformation,
\begin{equation}
	\label{eq:UnitaryMatrix}
	\hat{U}=
	\begin{pmatrix}
		1 & 0 & 0 \\
		0 & \dfrac{1}{\sqrt{2}} & \dfrac{i}{\sqrt{2}} \\
		0 & \dfrac{1}{\sqrt{2}} & -\dfrac{i}{\sqrt{2}}
	\end{pmatrix}
	\otimes I_{2\times 2}.
\end{equation}
Namely, the new basis set is
\begin{equation}
	\label{eq:OurBasis}
	\phi^{\dag} = (d^{\dag}_{0\uparrow},\ d^{\dag}_{0\downarrow},\ d^{\dag}_{1\uparrow},\ d^{\dag}_{1\downarrow},\ d^{\dag}_{2\uparrow},\ d^{\dag}_{2\downarrow})^{T},
\end{equation}
where $d^{\dag}_{0\sigma} = d^{\dag}_{z^2\sigma}$, $d^{\dag}_{1\sigma}$ = $(d^{\dag}_{x^2-y^2\sigma} + id^{\dag}_{xy\sigma})/\sqrt{2}$, and $d^{\dag}_{2\sigma}$ = $(d^{\dag}_{x^2-y^2\sigma} - id^{\dag}_{xy\sigma})/\sqrt{2}$ ($\sigma = \uparrow, \downarrow$).

We employ the standard $T$-matrix formalism to calculate QPI patterns.
Within this framework, the QPI signal is can be written as
\begin{equation}
	\label{eq:QPIsignal}
	\rho(\bm{q}, E) \propto\sum_{\bm{k}, i, f}\braket{\phi_{i}(\bm{k})|T(\bm{k}, \bm{q}, E)|\phi_{f}(\bm{k}+\bm{q})}\braket{\phi_{f}(\bm{k}+\bm{q})|\phi_{i}(\bm{k})}
\end{equation}
Here, $T(\bm{k}, \bm{q}, E)|$ is the $T$-matrix, $\ket{\phi_{i}(\bm{k})}$ and $\ket{\phi_{f}(\bm{k})}$ are the  eigenvectors of initial and final states, respectively.
The last term of Eq. \ref{eq:QPIsignal} $\braket{\phi_{f}(\bm{k}+\bm{q})|\phi_{i}(\bm{k})}$ represents the orthogonality between the initial and final states.
In this work, we focus on the effect of the spin and orbital selection rules originating from this last term of Eq. \ref{eq:QPIsignal}.
We assume spin- and orbital-preserving scatterings.
For all the our calculations, the energy broadening is set to 1~meV and the scattering potential is set to 0.1~eV for all orbital and spin channels.
The so-called setpoint effect~\cite{Kohsaka2007Science} and the lock-in broadening (20~mV$_\textrm{rms}$) are taken into account to compare with the experimental results.
Some more details of QPI simulations and comparison with experimental results are described in Ref. \onlinecite{Kohsaka2017PRB}.

For the full simulation including both the orbital and spin DOFs, the Green's function written with the basis set $\phi^\dag$ is used.
For the simulations where the orbital and/or spin DOF is hypothetically suppressed, the Green's functions written in a subspace of $\phi^\dag$ are used.
For the simulations including only the spin DOF, the Green's function of each orbital $G_i$ ($i = 0, 1, 2$) is written with $\phi^\dag_i = (d_{i\uparrow}^\dag,\ d_{i\downarrow}^\dag)^{T}$.
Similarly, for the simulations including only the orbital DOF, the Green's function of each spin $G_\sigma$ ($\sigma = \uparrow, \downarrow$) is written with $\phi^\dag_\sigma = (d_{0\sigma}^\dag,\ d_{1\sigma}^\dag,\ d_{2\sigma}^\dag)^{T}$.
To suppress both of the spin and orbital DOFs, the Green's function $G_{i\sigma}$ is written with $\phi_{i\sigma}^\dag = (d_{i\sigma}^\dag)$.
To calculation QPI patterns with DOF suppression, the sum of each Green's function, $\sum_iG_i$, $\sum_\sigma G_\sigma$, or $\sum_{i,\sigma} G_{i\sigma}$ is used.

The results of the full simulations are shown in Figs.~6(i)-6(l).
Contrary to experimental results, the signals originated from the $\Gamma_1$ band are quite strong as indicated by the red arrows in Figs.~6(k) and 6(l).
This contradiction seems to stem from the spectral weight broadening due to the three-dimensionality (a large dispersion along the $k_{\mathrm{z}}$ direction) of the $\Gamma_1$ band in the bulk 3R-materials [Fig.~1(b)]\cite{Suzuki2014NNano}, which is not contained in our tight-binding model for a monolayer TMDC.

\begin{figure*}[t]
\begin{center}
\includegraphics[width=18cm]{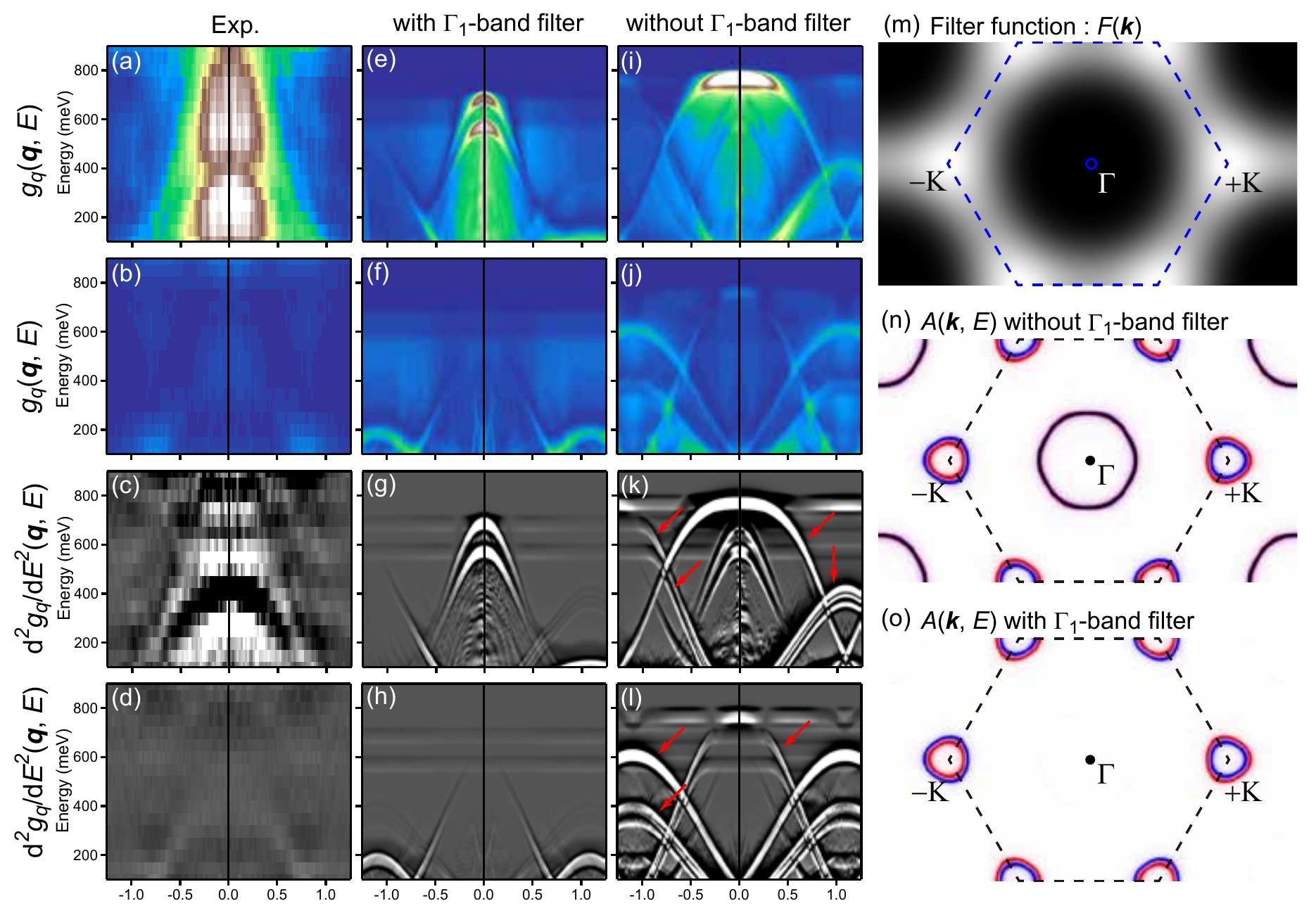}
\end{center}
\caption{
Effect of $\Gamma$ bands on simulated QPI patterns.
(a) and (b) Experimental data of energy dependent line profiles of $g_{q}(\bm{q}, E)$ along the red and white lines in Fig. 3(o), respectively.
(c) and (d) The second derivatives of (a) and (b), respectively.
(e) to (h) Simulations with the $\Gamma_1$-band filtering corresponding to (a) to (d), respectively. Both DOFs are included.
(i) to (l) Same as (e) to (h) but without the $\Gamma_1$-band filtering.
Red arrows in (k) and (l) represent the branches associated with the $\Gamma_1$-band.
(m) $\Gamma_1$-band filtering function $F(\bm{k})$ with a decay length of $k_{dec} = 0.6 \AA^{-1}$.
(n) and (o) Spin-resolved spectral functions $A(\bm{k}, E)$ at $E = $ 400 meV without and with the filtering function, respectively.
}
\end{figure*}

To include this broadening effect in our simulations, we multiply the bare Green's functions by a filtering function [$F(\bm{k})$] defined as follows;
\begin{eqnarray}
F(\bm{k}) & = & (1-f(\bm{k}))^4,\nonumber\\
f(\bm{k}) &= & (f'(\bm{k}) - f'_{\mathrm{min}})/(f'_{\mathrm{max}} - f'_{\mathrm{min}}), \nonumber\\
f'(\bm{k}) & = & \sum_{n=0}^{6} \frac{1}{2\pi w}\exp\left({-\frac{(k_x - \Gamma _{xn})^2+(k_y - \Gamma _{yn})^2}{2 w^2}}\right),\nonumber\\
\end{eqnarray}
where $(\Gamma _{xn}, \Gamma _{yn})$ is the position of $n$-th $\Gamma$ point ($\Gamma _0$ is center of 1st BZ, $\Gamma _{1 \sim 6}$ denote six $\Gamma$ points surrounding $\Gamma _0$), and $w$ is a half-width at half maximum of Gaussian distribution ($w$ = 0.6 $\rm{\AA}^{-1}$ in this work).
The resulting filtering function and the effect of this $\Gamma_1$-band filtering are summarized in Fig.~6.
Figures~6(e)-6(h) represent the simulated results with this $\Gamma_1$-band filtering as experimentally observed.
The signatures originated from the valley bands at $\pm$K are clearer than those in the simulations without this filtering.
Note that the conclusion in this paper does not change even if we do not use this $\Gamma_1$-band filtering.

\bibliography{3RNbS2_Refs}

\end{document}